\documentclass[twocolumn,secnumarabic,amssymb, nobibnotes, aps, prb, superscriptaddress]{revtex4-2}
\usepackage{lipsum}
\usepackage{titlesec}
\titlespacing\section{0pt}{12pt plus 4pt minus 4pt}{1pt plus 20pt minus 2pt}
\usepackage{xcolor}
\usepackage{amsmath}
\usepackage{comment}
\usepackage{physics}
\usepackage{afterpage}
\usepackage{placeins}
\usepackage{graphicx}
\usepackage{float}
\usepackage{booktabs}
\usepackage{multirow}
\usepackage{array}
\usepackage{setspace}
\graphicspath{{Figs/}}
\usepackage{siunitx}
\usepackage{hhline}
\usepackage{xfrac}
\usepackage{float,graphicx}
\usepackage{mathtools}
\usepackage{listings}
\usepackage{amssymb}
\usepackage{titlesec}
\usepackage{amsfonts}
\usepackage[version=4]{mhchem}

\usepackage{epstopdf}
\catcode`@11
\def\seceqaa{\@addtoreset{equation}{section}
\def\theequation{A\arabic{equation}}}
\def\seceqbb{\@addtoreset{equation}{section}
\def\theequation{B\arabic{equation}}}
\def\seceqcc{\@addtoreset{equation}{section}
\def\theequation{C\arabic{equation}}}
\def\seceqdd{\@addtoreset{equation}{section}
\def\theequation{D\arabic{equation}}}
\def\seceqee{\@addtoreset{equation}{section}
\def\theequation{E\arabic{equation}}}
\def\seceqff{\@addtoreset{equation}{section}
\def\theequation{F\arabic{equation}}}
\def\seceqgg{\@addtoreset{equation}{section}
\def\theequation{G\arabic{equation}}}
\def\seceqhh{\@addtoreset{equation}{section}
\def\theequation{H\arabic{equation}}}
\catcode`@11

\begin{document}

\title{Band splitting induced Berry flux and intrinsic anomalous Hall conductivity in NiCoMnGa quaternary Heusler compound} 

\author{Gaurav K. Shukla}
\affiliation{School of Materials Science and Technology, Indian Institute of Technology (Banaras Hindu University), Varanasi 221005, India}

\author{Jyotirmay Sau}
\affiliation{S. N. Bose National Centre for Basic Sciences, Kolkata 700098, West Bengal, India}

\author{Vishal Kumar}
\affiliation{School of Materials Science and Technology, Indian Institute of Technology (Banaras Hindu University), Varanasi 221005, India}

\author{Manoranjan Kumar}
\affiliation{S. N. Bose National Centre for Basic Sciences, Kolkata 700098, West Bengal, India}

\author{Sanjay Singh*}
\affiliation{School of Materials Science and Technology, Indian Institute of Technology (Banaras Hindu University), Varanasi 221005, India}

%\date{today}

\begin{abstract}
 The anomalous transport properties of Heusler compounds become a hotspot of research in recent years due to their unique band structure and possible application in spintronics. In this paper, we report the anomalous Hall effect in polycrystalline NiCoMnGa quaternary Heusler compound by experimental means and theoretical calculations. The experimental anomalous Hall conductivity (AHC) was found at about 256 $S/cm$ at 10K  with an intrinsic contribution of $\sim$ 121 $S/cm$. The analysis of Hall data reveals the presence of both extrinsic and intrinsic contributions in AHE. Our theoretical calculations show that a pair of spin-orbit coupled band formed by the band splitting due to spin-orbit interaction (SOI)
  at the Fermi level produces a finite Berry flux in the system that provides the intrinsic AHC about 100 $S/cm$, which is in good agreement with the experiment.
\end{abstract}

\maketitle
\section{INTRODUCTION}
%\vspace*{-3mm}
  The conventional Hall effect describes the phenomenon of transverse deflection of moving charges in a current carrying conductor placed in a magnetic field and this effect leads to a voltage difference perpendicular to both direction of motion of charges and the magnetic field \cite{nagaosa2006anomalous}. However, in ferromagnetic materials an additional large transverse voltage drop occurs in comparison to the normal conductors and this phenomena is known as anomalous Hall effect (AHE) \cite{nagaosa2010anomalous,tian2009proper,yue2017towards,manna2018heusler}. %The AHE was observed by E. Hall himself but the explanation of AHE was missing for a long decades. 
 So far, two kinds of mechanisms have been proposed to understand the origin of AHE; intrinsic and extrinsic mechanisms \cite{smit1955spontaneous,smit1958spontaneous,karplus1954hall,sidejump}.

In the intrinsic mechanism, the interband mixing along with spin-orbit interaction (SOI) results in an anomalous velocity of electrons perpendicular to the electric field direction \cite{karplus1954hall}. Subsequently, this anomalous velocity was reformulated in terms of the Berry phase and Berry curvature of Bloch bands \cite{sundaram1999wave, xiao2010berry,manna2018heusler}. 
%The Berry phase is the geometrical phase accumulated by the quantum mechanical system in a configuration space, when it ends the adiabatic motion at the start point along a closed path \cite{xiao2010berry}. 
Berry curvature is a pseudo-magnetic field in momentum space, which acts just like the magnetic field in electrodynamics \cite{xiao2010berry,manna2018heusler,stejskal2022flow}. 
%In real materials the Berry curvature is finite, when two or more bands approach to each other energetically and hybridize \cite{stejskal2022flow}. 
%The symmetry of the crystal brings the spin bands closer and SOI allows them to hybridize \cite{stejskal2022flow}. 
The SOI has small energy in comparison to the exchange splitting energy of ferromagnets, however the small energy of SOI can split the band dispersion near to the Fermi energy and when the Fermi energy lies in the SOI energy gap, a non zero dissipationless Hall current arises due to non-vanishing total flux of Berry curvature \cite{MnAs,iron,nagaosa2006anomalous}. The anomalous Hall conductivity (AHC) due to Berry phase (intrinsic mechanism) does not depend on the longitudinal conductivity (\si{\sigma}\textsubscript{\tiny{\textit{xx}}}) \cite{prb1,prb2}. 

The external origin of AHE, which includes the skew scattering and side jump mechanisms are known as extrinsic mechanism.  Skew scattering is an asymmetric scattering of carriers by impurity potential, which introduces a momentum perpendicular to both the incident wave vector \textit{\textbf{k}} and magnetization M. AHC due to skew scattering is proportional to the \si{\sigma}\textsubscript{\tiny{\textit{xx}}} \cite{smit1955spontaneous}.
Side jump is a microscopic displacement of wave packet due to SOI coupled Bloch states under the influence of disorder potential and the AHC due to side jump is independent of \si{\sigma}\textsubscript{\tiny{\textit{xx}}} \cite{MnAs,sidejump}.

%The AHE is a fundamental transport phenomenon in magnetic conductors that arises due to the mutual communication of the spin-orbit interaction (SOI) and spin polarization of electric current in ferromagnets \cite{nagaosa2010anomalous,tian2009proper,yue2017towards,manna2018heusler}.
Currently, there is an immense interest in the AHE due to its potential applications in spintronics such as for magnetic sensors and memory devices \cite{ohno2000electric,PhysRevLett.104.106601,hao2017anomalous}. Several materials have been reported to have large intrinsic AHC due to the Berry curvature originating from their characteristic spin-orbit coupled band structure. For example, the intrinsic AHC in Fe (751 $S/cm$) \cite{iron} and MnAs ($\sim$ 10$^3$ $S/cm$) \cite{MnAs} have been reported due to the Berry curvature arising from a pair of spin-orbit coupled band near to the Fermi level. Co$_3$Sn$_2$S$_2$ \cite{liu2018giant}, Fe$_3$Sn \cite{chen2022large} and LiMn$_6$Sn$_6$ \cite{dong} show the large intrinsic AHC arising due to the finite Berry curvature associated with the band structure. Besides these materials, Heusler alloys \citep{wollmann2017heusler,graf2011simple,elphick2021heusler} show the exotic anomalous transport properties owing to the distinctive band structure due to the combined effect of SOI and broken time-reversal symmetry \cite{prb1,manna2018colossal,Xu15,belopolski2019discovery}. For example, Co$_2$MnAl \cite{li2020giant}, Co$_2$MnGa \cite{manna2018colossal}, Fe$_2$-based Heusler alloys \cite{mende2021large} show the large intrinsic AHE.

 Co$_2$MnGa is a full Heusler compound that crystallizes in the L2$_1$ structure with space group Fm${\bar3}$m, consists of 
 %four interpenetrating FCC sublattices in which two are formed by Co atoms and the other two sublattices are formed by the Mn and Ga atoms \cite{guin2019anomalous}. Without spin-orbit coupling (SOC), there are 
 three nodal lines near Fermi energy, derived from the three mirror symmetries present in the system in the absence of spin-orbit coupling (SOC)  \cite{guin2019anomalous}.  With the consideration of SOC, these nodal lines gap out according to the magnetization direction and create the large Berry curvature and intrinsic AHE in the system \cite{belopolski2019discovery,guin2019anomalous}.
In the present manuscript, we studied the AHE in NiCoMnGa quaternary Heusler compound, which can be obtained by replacing the one Co atom by its neighbouring Ni atom in Co$_2$MnGa full Heusler compound keeping the magnetic moments nearly close for both the compounds.
Experimentally, we found the value of AHC around 256 $S/cm$ at 10K with an intrinsic contribution of $\sim$ 121 $S/cm$. The theoretical calculations give the intrinsic AHC $\sim$ 100 $S/cm$, which is in good agreement with the experiment. The reduction of the mirror symmetries in the NiCoMnGa in comparison to the Co$_2$MnGa lead to the absence of nodal line, nevertheless the band splitting in presence of SOC at Fermi energy leads to the finite Berry curvature and intrinsic AHC in the system.
\section{METHODS}
Polycrystalline NiCoMnGa Heusler compound was synthesized by arc melting method in the environment of high pure argon atmosphere by taking 99.99 \% pure  constituent elements in the water cooled copper hearth. To reduce further contamination, a Ti piece was used as an oxygen getter. The sample was flipped several times and re-melted for homogeneous mixing. The energy dispersive x-ray
(EDX) analysis suggests a composition ratio of 1:1:1:1 within the standard deviation (3\%-5\%) of the EDX measurement. For the structural analysis x-ray diffraction (XRD) pattern of powder sample was recorded at room temperature using Rigaku made x-ray diffractometer with Cu-\textit{K}\textsubscript{$\alpha$} radiation. The magnetization measurements were performed using a vibrating sample magnetometer (VSM) attached with Physical Properties Measurement System (PPMS) from Quantum Design. The resistivity and Hall measurement were carried out on a rectangular piece of sample of dimension 4.34\,$\times\,$2.45$\times\,$0.66 mm$^3$ employing four-probe method  using Cryogen Free Measurement System (CFMS).
The electronic band structure and magnetic properties of the NiCoMnGa were calculated by employing density functional theory (DFT) using the Vienna ab initio simulation package (VASP) \cite{hafner2008ab}. The generalized gradient approximation (GGA) of Perdew-Burke-Ernzerhof (PBE) type was used for the exchange-correlation functional \cite{blochl1994projector}. 
\begin{figure}[t]
    \centering
    \includegraphics[width=0.5\textwidth]{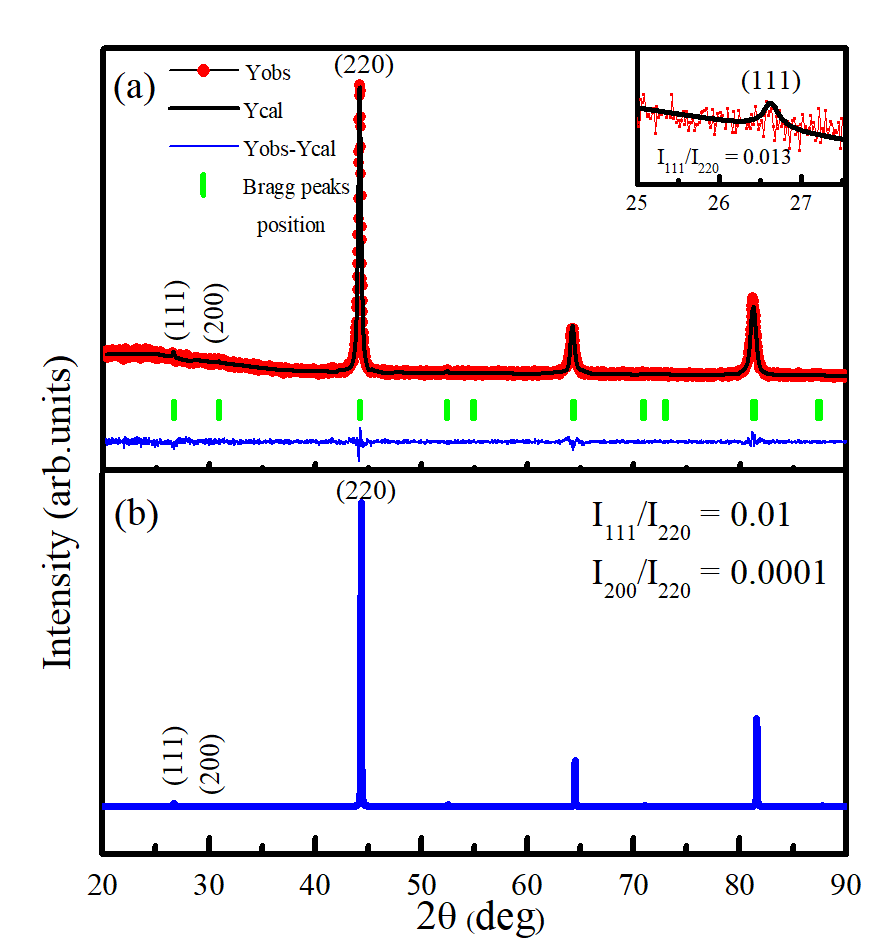}
    \caption{Rietveld refinement of the room temperature powder x-ray diffraction (XRD) data of NiCoMnGa system. The inset shows an enlarged view of XRD pattern around (111) superlattice reflection. (b) Simulated XRD pattern of NiCoMnGa system.}
    \label{Fig1}
\end{figure}
\begin{figure}[t]
\centering
\includegraphics[width=0.4\textwidth]{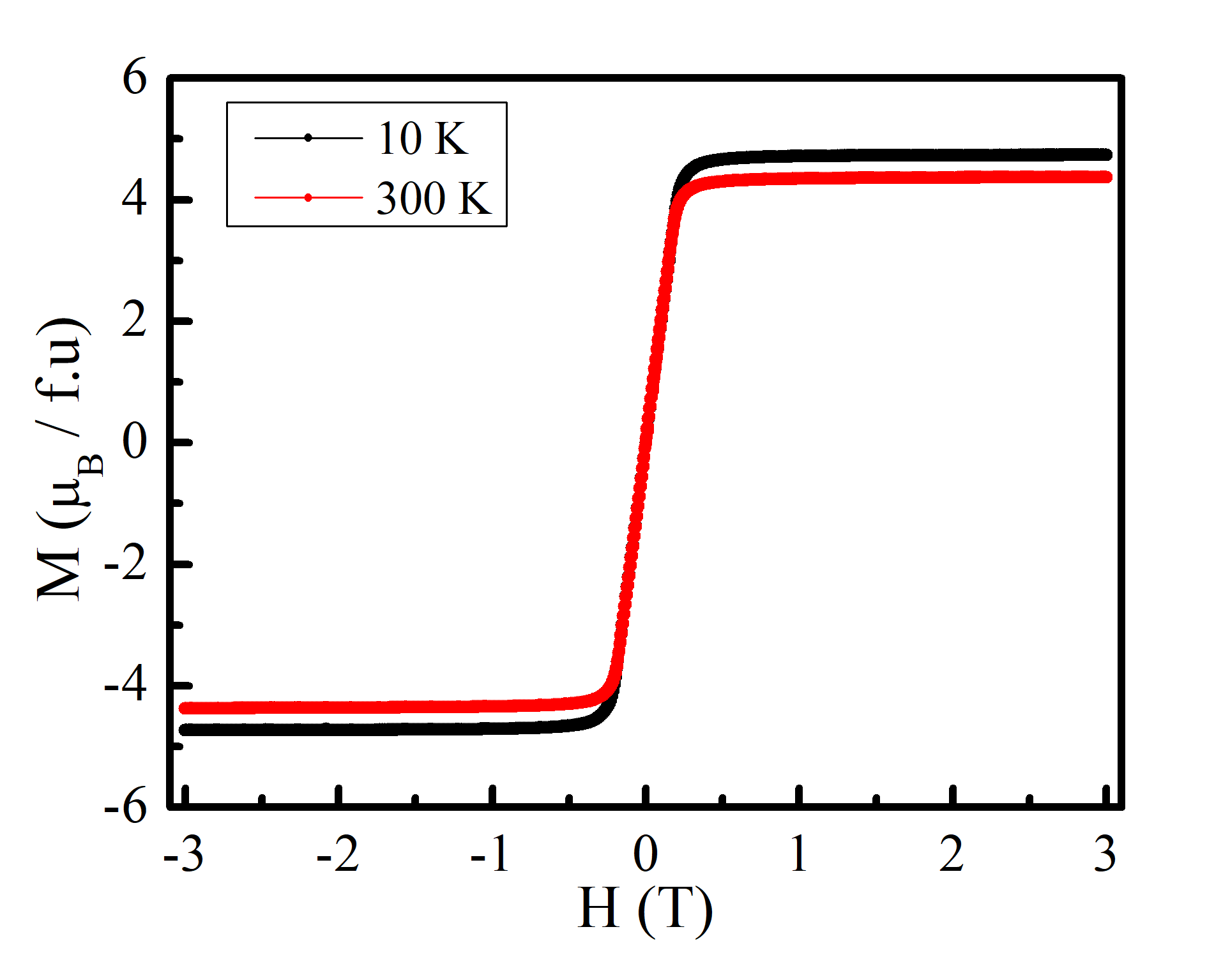}
\caption{Field-dependent magnetic isotherms at 10\,K and 300\,K.}
\label{Fig2}
\end{figure}
\begin{figure*}[t]
    \centering
    \includegraphics[width=1\textwidth]{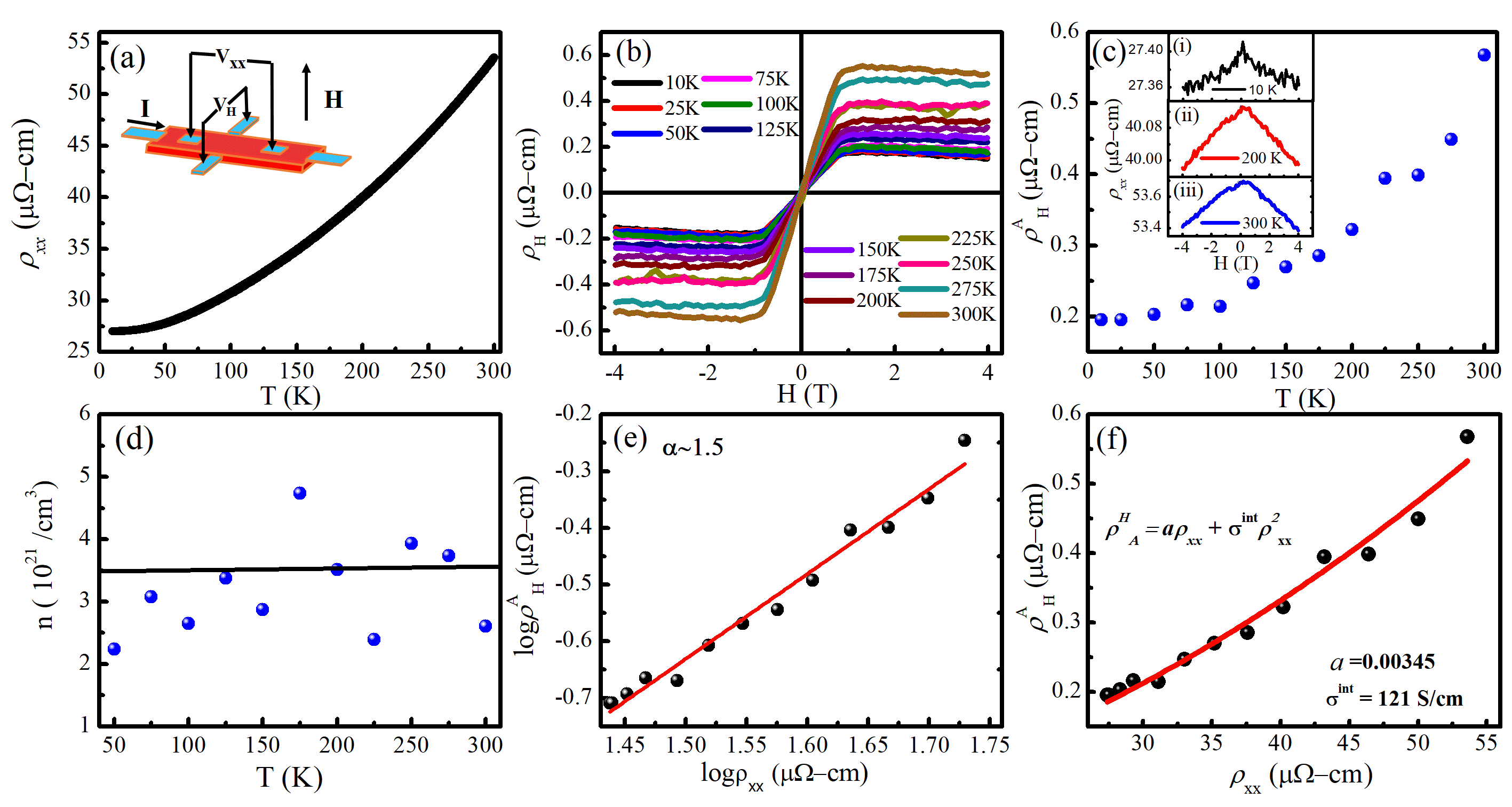}
    \caption{(a) Temperature-dependent longitudinal resistivity \si{\rho}\textsubscript{\tiny{\textit{xx}}}. Inset shows a schematic drawing of sample device used for Hall voltage V$_H$ and longitudinal voltage V$_{xx}$ measurements. (b) Field-dependent Hall resistivity $\si{\rho}\textsubscript{\tiny{\textit{H}}}$ at different temperatures. (c) Temperature variation of anomalous Hall resistivity $\si{\rho}^{A}_{\textsubscript{\tiny{\textit{H}}}}$. Inset shows the field-dependent \si{\rho}\textsubscript{\tiny{\textit{xx}}} at (i) 10\,K,\,(ii) 200\,K, and (iii)\, 300\,K. (d) Temperature-dependent carrier concentration \textit{n}. (e) Double logarithmic plot between $\si{\rho}^{A}_{\textsubscript{\tiny{\textit{H}}}}$ and \si{\rho}\textsubscript{\tiny{\textit{xx}}} (black balls) and the linear fitting is shown by a red line. (f) $\si{\rho}^{A}_{\textsubscript{\tiny{\textit{H}}}}$ and \si{\rho}\textsubscript{\tiny{\textit{xx}}} plot (black balls) and fitted curve using Eq.(5) is shown in red color .}
    \label{Fig3}
\end{figure*} 
\begin{figure}[t]
\centering
\includegraphics[width=0.5\textwidth]{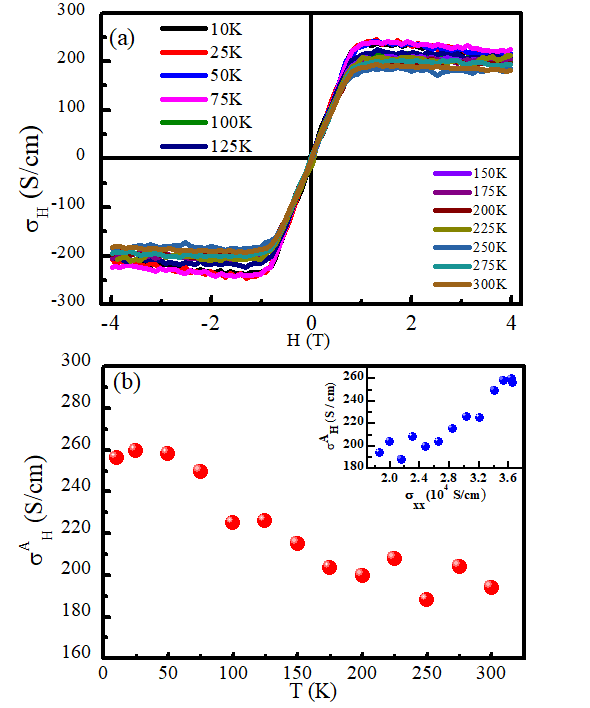}
\caption{(a) Field-dependent Hall conductivity \si{\sigma}\textsubscript{\tiny{\textit{H}}} at different temperatures. (b) Temperature variation of anomalous Hall conductivity (AHC). Inset shows the variation of AHC with longitudinal conductivity \si{\sigma}\textsubscript{\tiny{\textit{xx}}}.}
\label{Fig4}
\end{figure}
The kinetic energy cutoff of 520 eV was taken for the plane-wave basis. A 15 × 15 × 15 k-point mesh was used for the Brillouin zone (BZ) sampling and the Gaussian smearing method with a width of 0.1 eV was adopted for the Fermi surface broadening. 
%Both 
The cell parameter 
%and internal atomic positions were fully 
was relaxed until the forces on  all atoms were smaller than 0.01 eV/Å. The SOC was taken into account in all the calculations. To explore the non-trivial band topology and the intrinsic AHC, the tight binding Hamiltonian was constructed with the maximally localized Wannier functions using the Wannier90 code \cite{Pizzi_2020,marzari_97}. Based on the tight-binding Hamiltonian, the AHC and the Berry curvature were evaluated via the Kubo-formula approach \cite{Gradhand_2012}.

\section{RESULTS AND DISCUSSION}
\subsection{Structure and magnetization}
%\vspace*{-3mm}
%\subsection{Structural properties}\vspace*{-3mm}
The XRD pattern of NiCoMnGa was recorded at room temperature for structural analysis. The observed XRD pattern (red curve in Fig.\,\ref{Fig1}(a)) fairly shows the cubic structure of the sample. The quaternary Heusler alloys generally crystallize in the LiMgPdSn type structure (space group \textit{F$\bar{4}$3m})\cite{quaternary}. 
The Rietveld refinement of the room temperature XRD pattern of NiCoMnGa was carried out using FULLPROF software package \cite{FullPROF} in the space group of \textit{F$\bar{4}$3m} (space group no.216) and the  special Wyckoff's positions; 4d (0.75,0.75,0.75), 4c (0.25, 0.25,0.25,0.25), 4b (0.5, 0.5, 0.5), and 4a (0, 0, 0) were considered for Nickel (Ni), Cobalt(Co), Manganese (Mn) and Gallium (Ga) atoms, respectively.
The calculated XRD pattern (black curve) depicted in  Fig.\,\ref{Fig1}(a), shows that all the Bragg peaks are well-indexed, which confirms the single-phase (cubic) of the sample. The  refined lattice parameter was found 5.79 \AA, which matches well with the value reported in the literature \cite{vajiheh}. 
The presence of the (111) and (200) superlattice reflections generally mark the ordered structure of Heusler compounds \cite{prb1}. For NiCoMnGa the structure factors for (111), (200) and (220) reflections can be written as \cite{quaternary2} 
 \begin{equation}
     F_{111} = 4[(f_{Ga} - f_{Mn})-i (f_{Ni} - f_{Co})]
 \end{equation}
 \begin{equation}
     F_{200} = 4[(f_{Ga} + f_{Mn}) - (f_{Ni} + f_{Co})]
 \end{equation}
\begin{equation}
     F_{220} = 4[(f_{Ga} + f_{Mn}) + (f_{Ni} + f_{Co})]
 \end{equation}
 The negligible intensity of the (111) superlattice reflection survives only due to the difference of the atomic scattering factors (SFs) of Ga and Mn atoms as the difference between atomic SFs of Ni and Co atoms is negligible (both are the consecutive elements in periodic table) \cite{prb1,quaternary2}. The vanishing intensity of the (200) reflection can be understood from Eq.2. The intensity of fundamental reflection is due to the sum of all atomic SFs  of constituent elements (Eq.3). We simulated the XRD pattern of NiCoMnGa using POWDERCELL software \cite{kraus1996powder} depicted in the Fig.\,\ref{Fig1}(b). The intensity ratio of superlattice reflection (111) to the fundamental reflection (220) (\textit{i.e.} $\frac{I_{111}}{I_{220}}$) was found about 0.01 and 0.012 from the simulated and observed XRD patterns, which suggests the formation of ordered structure of NiCoMnGa Heusler compound. Inset of Fig.\,\ref{Fig1}(a) shows the enlarged view of XRD pattern around (111) superlattice reflection. Figure\,\ref{Fig2} shows the magnetic isotherms at the temperature 10\,K and 300 K. The magnetic moment was found 4.7 $\mu_B$/f.u at 10\,K. The magnitude of the  observed magnetic moment is close to the value reported in literature\cite{vajiheh}. 
\subsection{Transport measurements}
Figure\,\ref{Fig3}(a) shows the measured temperature dependence of longitudinal resistivity (\si{\rho}\textsubscript{\tiny{\textit{xx}}}). The temperature variation of \si{\rho}\textsubscript{\tiny{\textit{xx}}} indicates the metallic conduction with \si{\sigma}\textsubscript{\tiny{\textit{xx}}} about 1.88$\times$10$^4$ $S/cm$ at 300\,K. The residual resistance ratio (RRR= $\rho\textsubscript{xx}(300 K)$/$\rho\textsubscript{xx} (10 K)$),
which quantifies the degree of disorder, is found around 2, which suggests a clean sample of NiCoMnGa \cite{prb1}. Now, afterward investigating the phase purity, magnetization and resistivity of the sample, we will discuss the outcomes of Hall measurement.
%\subsection{Anomalous Hall}
Inset of Fig.\,\ref{Fig3}(a) shows the schematic diagram of sample device used for longitudinal voltage (V$_{xx}$) and Hall voltage (V$_H$) measurements.
In general the Hall resistivity (\si{\rho}\textsubscript{\tiny{\textit{H}}}) in magnetic materials is sum of two parts \cite{prb1,nagaosa2006anomalous}; 
\begin{equation}
 \si{\rho}\textsubscript{\tiny{\textit{H}}}=\si{\rho}^{0}_{\textsubscript{\tiny{\textit{H}}}}+\si{\rho}^{A}_{\textsubscript{\tiny{\textit{H}}}}= R_0H +R_sM  
\end{equation}
$\si{\rho}^{0}_{\textsubscript{\tiny{\textit{H}}}}$ and $\si{\rho}^{A}_{\textsubscript{\tiny{\textit{H}}}}$ are the ordinary and anomalous Hall resistivity, respectively.
 R$_0$, R$_s$, H and M are the ordinary Hall coefficient, anomalous Hall coefficients, applied external magnetic field and spontaneous magnetization of material, respectively.  R$_0$ which depends on the type of charge carriers and their density is the inverse of the product of carrier concentration (n) and electronic charge (e) \cite{wang2018large}. The Hall resistivity in ferromagnetic materials dominates by AHE at the lower field and the role of ordinary Hall effect usually appears in the higher field region \cite{prb1}.
With the linear fitting of the high field \si{\rho}\textsubscript{\tiny{\textit{H}}} curve, slope and intercept on the y-axis give R$_0$ and $\si{\rho}^{A}_{\textsubscript{\tiny{\textit{H}}}}$, respectively. $\si{\rho}^{A}_{\textsubscript{\tiny{\textit{H}}}}$ is alike to Hall response due to magnetization of material in absence of an external magnetic field. Hall measurement outcomes are summarized in Fig.\,\ref{Fig3} (b)-(f) and Fig.\,\ref{Fig4}\,(a)-(b). Figure\,\ref{Fig3}(b) shows the field dependent $\si{\rho}\textsubscript{\tiny{\textit{H}}}$ isotherms  up to field of 4\,T. The \si{\rho}\textsubscript{\tiny{\textit{H}}} curves show a sharp jump at low field and change linearly in the high field regime that signifies an AHE in the present material. 
Figure\,\ref{Fig3}(c) displays the variation of extracted  $\si{\rho}^{A}_{\textsubscript{\tiny{\textit{H}}}}$ with temperature. $\si{\rho}^{A}_{\textsubscript{\tiny{\textit{\textit{H}}}}}$ increases non-linearly with temperature and achieves the maximum value of about 0.56 $\mu\Omega$-cm at room temperature. We also measured the field-dependent $\si{\rho}\textsubscript{\tiny{\textit{xx}}}$ at fixed temperatures. The $\si{\rho}\textsubscript{\tiny{\textit{xx}}}$ does not changes significantly with the magnetic field at a particular temperature as shown in the inset of Fig.\,\ref{Fig3}(c) for (i) 10\,K,\,(ii) 200\,K,\,and (iii) 300\,K. The charge carrier density (\textit{n}) was calculated using relation R$_0$=$\frac{1}{ne}$ \cite{wang2018large} and temperature variation of \textit{n} is depicted in Fig.\,\ref{Fig3}(d). The magnitude of \textit{n} is a little scattered with temperature and estimated about 2.5$\times$10$^{21}\,$cm$^{-3}$ at 300 \,K. In order to examine the origin of AHE, we have plotted the $\si{\rho}^{A}_{\textsubscript{\tiny{\textit{H}}}}$ versus \si{\rho}\textsubscript{\tiny{\textit{xx}}} on a double logarithmic scale as shown in Fig.\,\ref{Fig3}(e). A linear fitting was done to determine the exponent $\alpha$ according to the formula $\si{\rho}^{A}_{\textsubscript{\tiny{\textit{\textit{H}}}}}$ $\propto$ $\si{\rho}^\alpha_{\textsubscript{\tiny{\textit{xx}}}}$ \cite{roy2020anomalous,wang2018large}. If $\alpha $ = 1, the origin of AHE is due to skew scattering and if $\alpha $ = 2, the origin of AHE is assigned to the intrinsic and side jump mechanisms \cite{nagaosa2010anomalous,prb1,roy2020anomalous}. In this way, we found the exponent $\alpha = 1.50$, which primarily specifies that the extrinsic and intrinsic  mechanisms are involved in the AHE.
 To disentangle the intrinsic and extrinsic contributions linked with AHE, we have used the following equation, which accounts the phonon contribution in the skew scattering as suggested for alloys system \cite{prb1,PhysRevB.100.054445, wang2018large, PhysRevB.96.144426}
\begin{equation}
  \si{\rho}^{A}_{\textsubscript{\tiny{\textit{H}}}} = a\si{\rho}\textsubscript{\tiny{\textit{xx}}} + \si{\sigma}^{int}\si{\rho}{^2_{\textsubscript{\tiny{\textit{xx}}}}}.
\end{equation}
Here a is the parameter related to the skew scattering and the notation \si{\sigma}$^{int}$ is used for intrinsic AHC. Here we have assumed the ratio of SOI energy ($\epsilon\textsubscript{SO}$) to the Fermi energy (E\textsubscript{F}) \textit{i.e.} $\epsilon\textsubscript{SO}$/E\textsubscript{F} is in order of 10$^{-3}$ to 10$^{-2}$, which leads the suppression of side jump contribution in comparison to intrinsic AHC as observed for other metallic ferromagnets \cite{prb1,roy2020anomalous}.
Fig.\,\ref{Fig3}(f) shows the \si{\rho}$^{A}_{\textsubscript{\tiny{\textit{H}}}}$ versus \si{\rho}\textsubscript{\tiny{\textit{xx}}} plot (black balls) and the fitting was employed using Eq.(5) as shown by a red curve. From fitting, the parameter $a$ and \si{\sigma}$^{int}$ comes out to be 0.0034 and 121 $S/cm$, respectively. The change in $\si{\sigma}^{A}_{\textsubscript{\tiny{\textit{H}}}}$ with temperature and/or longitudinal conductivity (\si{\sigma}\textsubscript{\tiny{\textit{xx}}}) also holds the information about the involved mechanism in the AHE. Hall conductivity (\si{\sigma}\textsubscript{\tiny{\textit{H}}}) was calculated  using equation \cite{guin2019anomalous,prb1,prb2}
\begin{equation}
    \si{\sigma}\textsubscript{\tiny{\textit{H}}}= \frac{\si{\rho}\textsubscript{\tiny{\textit{H}}}}{(\si{\rho}{^2_\textsubscript{\tiny{\textit{xx}}}}+\si{\rho}{^2_\textsubscript{\tiny{\textit{H}}}})}
\end{equation}
\begin{figure*}[htbp]
\centering
\includegraphics[width=0.8\textwidth]{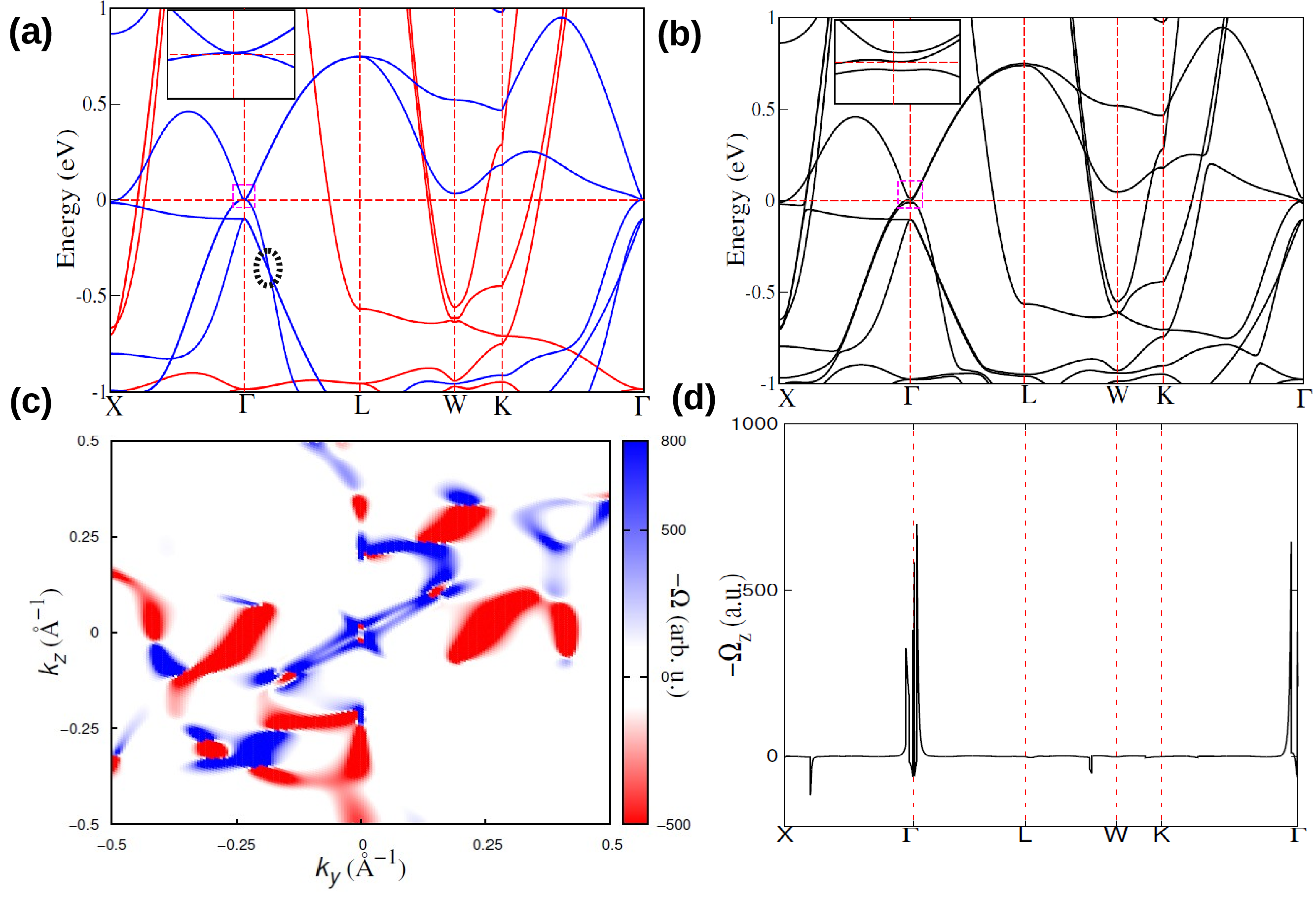}
\caption{(a) The band structure of NiCoMnGa in the absence of SOC, where the red and blue colors represent the spin-up and spin-down bands, respectively. (b) The band structure of NiCoMnGa in the presence of SOC. The inset shows the splitting of the bands near the Fermi energy E$_F$. (c) Berry curvature distribution in $\textit{k}_y$-$\textit{k}_z$ plane at $\textit{k}_x=0$. (d) Berry curvature along the high-symmetry path.}
\label{Fig5}
\end{figure*}
 Figure\,\ref{Fig4} (a) shows the field-dependent Hall conductivity curves at different temperatures. The value of AHC at a particular temperature is calculated by zero-field extrapolation of the high field Hall conductivity data with the y-axis. We found the AHC around 256 $S/cm$ at 10\,K. This value is nearly twice of the magnitude of the intrinsic AHC, which shows the presence of equal contribution of intrinsic and extrinsic AHC at 10\,K. The AHC decreases with increasing the temperature and reduces to 194 $S/cm$ at 300 \,K as shown in Fig.\,\ref{Fig4}(b). The variation of AHC with \si{\sigma}\textsubscript{\tiny{\textit{xx}}} is shown in the inset of Fig.\,\ref{Fig4}(b). The decreasing (increasing) value of AHC with temperature ( $\si{\sigma}\textsubscript{\tiny{\textit{xx}}}$) is due to the skew scattering contribution in AHC rather than the temperature variation of spontaneous magnetization of sample as the temperature has a little effect on the magnetization of the present system (Fig.\,\ref{Fig2}). It is worthwhile to mention here that intrinsic AHE is expected to dominate in the overall behaviour of AHE, when the longitudinal conductivity of sample lies in the good metallic regime \textit{i.e.} \si{\sigma}\textsubscript{\tiny{\textit{xx}}} is an order of 10$^4$-10$^6$ $S/cm$ 
 \cite{prb1,prb2,roy2020anomalous,nagaosa2006anomalous,jinhu}. Our system shows the deviation from this criterion as the extrinsic AHE has significant contribution in the AHE despite the \si{\sigma}\textsubscript{\tiny{\textit{xx}}} is an order of 10$^4$ $S/cm$.
 The AHC due to skew scattering can be given as \cite{int&ext}
 \begin{equation}
    \si{\sigma}^{skew}_{\textsubscript{\tiny{\textit{H}}}} = \si{\sigma}\textsubscript{\tiny{\textit{xx}}}S = \frac{2e^2}{ha}\frac{E_F
{\tau}}{\hslash}S
 \end{equation}
  where  h, a, E$_F$, {$\tau$} and S are the Planck constant, lattice parameter, Fermi energy, mean free path of electron and skewness factor, respectively. The skewness factor is S $\sim$ {$\epsilon_{SO}$}v$_{imp}$/W$^2$, where W is the bandwidth and v$_{imp}$ is impurity potential \cite{int&ext}. In the superclean limit where the $\frac{\hslash}{\tau} \textless\textless $ $\epsilon_{SO}$, the skew scattering dominates, however in the case of $\frac{\hslash}{\tau}$ \textless $\epsilon_{SO}$, the skew scattering contribution decreases. From Eq.7, it is clear that the skew scattering contribution rapidly decays with increasing the $\frac{\hslash}{\tau}$. Henceforth, the value of AHC due to skew scattering depends on both the strength of SOI energy and the \si{\sigma}\textsubscript{\tiny{\textit{xx}}}. Our experiment also suggests that the lineup of the origin of AHE based solely on the longitudinal conductivity is a rough estimation and need not to be strictly valid. Since the intrinsic AHE depends on the Berry curvature of the system, which may have different origin such as presence of gapped nodal line, Weyl point or interband mixing along a certain high symmetry path. Therefore, to understand the origin of Berry curvature in present system, we performed the first principles calculation.
\vspace{-1mm}
\subsection{First principles calculation}
 The relaxed lattice parameter of NiCoMnGa was found 5.782 \AA, which is consistent with the experiment.
Our first principles calculation for magnetic moment suggests that Co and Mn have a large magnetic moment with $\mu_{Co}$=1.188 $\mu_B$/f.u and $\mu_{Mn}$=3.246 $\mu_B$/f.u  respectively, whereas Ga has a small vanishing magnetic moment and Ni have a moment  $\mu_{Ni}$=0.563 $\mu_B$/f.u. The total magnetic moment per formula unit is 4.993 $\mu_B$, aligned along the (001) direction, which is consistent with Slater Pauling rule for Heusler alloys \cite{ST}. To calculate the Berry curvature and the intrinsic AHE in the present system, the tight-binding Hamiltonian was constructed with the maximally localized Wannier functions \citep{Pizzi_2020, marzari_97}. Based on the tight-binding Hamiltonian, the AHC and the Berry curvature were evaluated via the Kubo-formula \cite{Gradhand_2012} approach in the linear response scheme as follows;
\begin{equation}
\sigma_{\alpha\beta} = -{\frac{e^2}{\hbar} \sum_{n}\int\frac{d^{3}K}{(2\pi)^3}\Omega^n_{\alpha\beta}f_n}  \end{equation}where Berry curvature $\Omega$ can be written as sum over eigen state using Kubo formula\cite{xiao2010berry}
 \begin{eqnarray}
\Omega^n_{\alpha \beta} = i \sum_{n \neq n'} \frac{{\langle n|\frac{\partial H}{\partial R^\alpha}|n'\rangle} {\langle n'|\frac{\partial H}{\partial R^\beta}|n \rangle}-(\alpha\xleftrightarrow{}\beta)}{(\epsilon_n - \epsilon_n')^2}
\end{eqnarray}
 Here $\ket{n}$ and $\epsilon_n$ are the energy eigenstate and  eigenvalue of Hamiltonian H, respectively. f$_n$ is the Fermi distribution function.
 The intrinsic AHE can be analyzed by exploring the electronic band structure of the NiCoMnGa. Two major features in electronic band structure make this system important; first, a twofold degenerate band form a triple point crossing with a non-degenerate band along high symmetry direction $\Gamma$-L in the absence of SOC as shown in Fig.\,\ref{Fig5}a by a black dotted circle. This doubly degenerate band splits in the presence of SOC and also lifts the triple point degeneracy.  
 %The triple point degeneracy is shown in the circle in Fig.\ref{Fig6}(a). 
 This degeneracy may arise due to a  symmetry of C$_{3v}$ along $\Gamma$-L high symmetry direction, whose elements are threefold rotation (C$_3$) and a $\sigma_v$ mirror plane \cite{TR}. 
 Second feature is a pair of degenerate minority spin band at $\Gamma$ point in absence of SOC, however, in the presence of SOC these bands split into an occupied and an unoccupied bands  separated by a small energy gap in a small \textit{k} interval near $\Gamma$ point as shown in Fig.\,\ref{Fig5}b. In Figure\,\ref{Fig5}d, we calculated the Berry curvature $\Omega^z_{xy}$ in the presence of SOC along same symmetry direction and $\Omega^z_{xy}$ has large value near the $\Gamma$ point. This large value of Berry curvature can be explained in terms of splitting of degenerate band, which opens a small energy gap and creates intrinsic AHC in the system.
 \begin{figure}[htbp]
\centering
\includegraphics[width=0.4\textwidth]{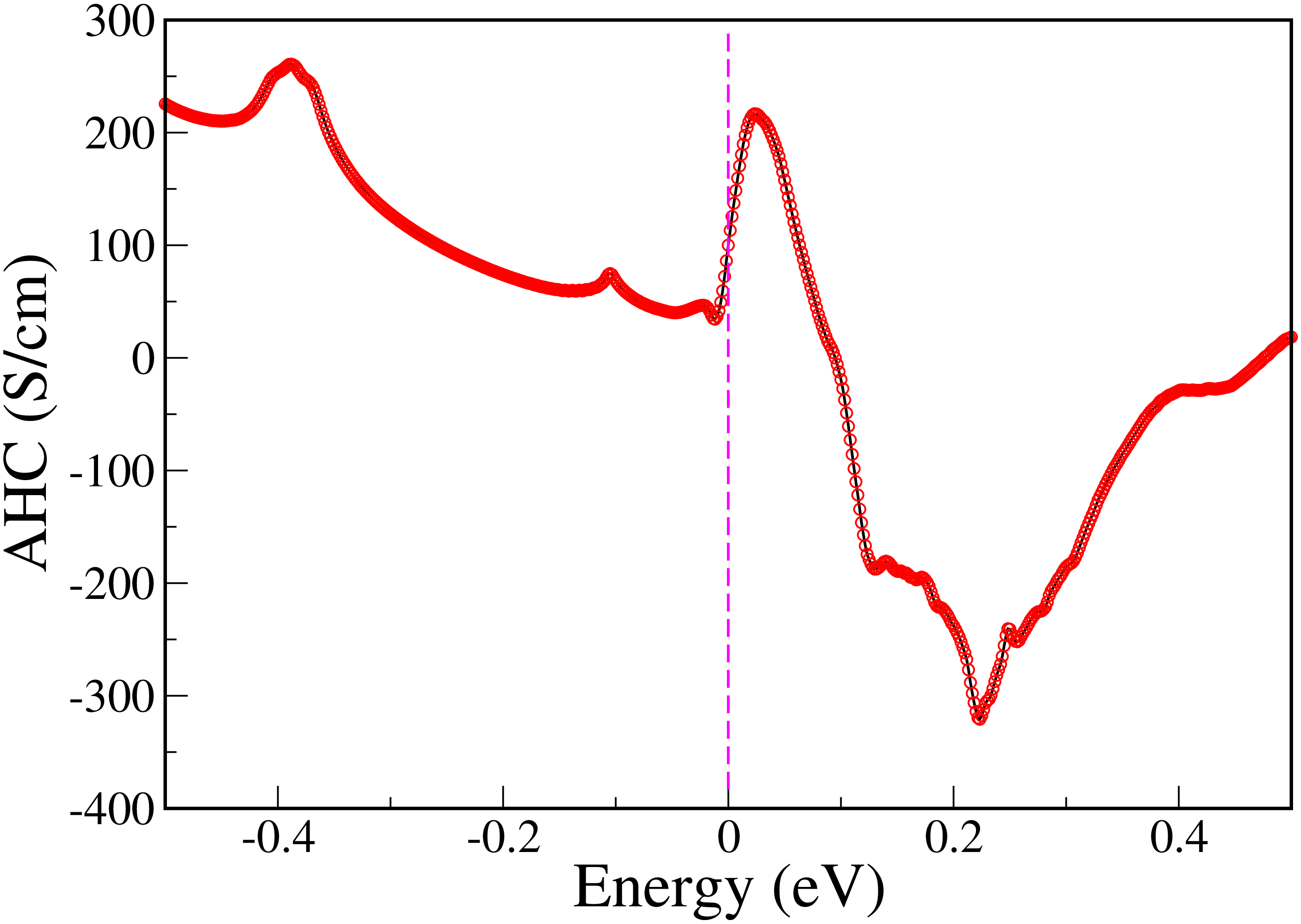}
\caption{Fermi energy scan of the anomalous Hall conductivity.}
\label{Fig6}
\end{figure}
 Owing to the nearly degenerate conduction and valance band at $\Gamma$ point, the Berry curvature is large at this point; therefore an intrinsic AHE is expected in the present compound. In the Fig.\,\ref{Fig5}(c), the color plot of Berry curvature is shown in $k_x$ =0 plane of the first Brillouin zone, which shows positive and negative distribution of Berry curvature in the plane. Following the symmetry of NiCoMnGa compound, if the magnetization is taken along the z-axis, $\Omega^z_{xy}$ is the only survival component of the Berry curvature. We found the intrinsic AHC due to finite $\Omega^z_{xy}$ about 100 $S/cm$ at Fermi level, which is consistent with the experiment (121 $S/cm$). 
 We also showed the dependency of AHC as a function of shift in Fermi energy in Fig.\,\ref{Fig6}, which gives an idea about contribution of various valance bands by shifting the Fermi energy. The large peak below the Fermi level (-0.38 eV) is due to large Berry curvature at the triple point, which is contributing as an occupied valence band to AHC.  We observed the decreasing trend of AHC from -0.38 eV and it again increases at Fermi energy due to a large positive Berry curvature shown by blue color in Fig.\,\ref{Fig5}(c). 

\section{CONCLUSIONS}
 We have experimentally measured the AHC in NiCoMnGa quaternary Heusler compound and theoretically calculated the intrinsic part of AHC due to Berry curvature of the dispersion bands. The extrinsic and intrinsic mechanisms both contribute equally in the AHC of the present system. We found a good agreement between the extracted intrinsic AHC experimentally and the theoretically calculated AHC. The reduction of the number of mirror symmetries in the NiCoMnGa in comparison to the Co$_2$MnGa lead to the absence of nodal line, nevertheless the band splitting in the presence of SOC at Fermi energy leads to the finite Berry curvature and intrinsic AHC in the system. The presence of signification contribution of extrinsic mechanism in AHE, despite the longitudinal conductivity is an order of 10$^4$ $S/cm$, suggests that the relation of the origin of AHE solely with the longitudinal conductivity may not be strictly valid. 
\section*{ACKNOWLEDGMENT}
We gratefully acknowledge IIT BHU for experimental support. SS thanks Science and Engineering Research Board of India for financial support through the award of Ramanujan Fellowship (grant no: SB/S2/RJN-015/2017) and UGC-DAE CSR, Indore for financial support through “CRS” Scheme. G.K.S. acknowledges the DST-INSPIRE scheme for support through a fellowship. M.K thanks DST for funding
through grant no.CRG/2020/000754. J.S. thanks University Grants Commission (UGC) for Ph.D. fellowship. V.K. thanks CSIR, New Delhi for financial support.

*ssingh.mst@itbhu.ac.in
\end{document}